\documentstyle[12pt]{article}
\begin{document}

\title{Effective Hamiltonians with Relativistic Corrections\\
{\large \it II) Application to Compton Scattering by a Proton}}
\author{S.~Scherer, G.~I.~Poulis and H.~W.~Fearing\\
TRIUMF, Vancouver, British Columbia, Canada V6T 2A3}
\date{}
\maketitle
\begin{abstract}
We discuss two different methods of obtaining ``effective $2 \times 2$
Hamiltonians'' of the electromagnetic interaction which include relativistic
corrections.
One is the standard Foldy--Wouthuysen transformation which we compare
with the Hamiltonian obtained from a direct reduction of the matrix element
of the interaction Hamiltonian between positive--energy solutions of
the free Dirac equation.
The two approaches are applied to Compton scattering by a proton for
which a low--energy theorem exists.
It is found that the Foldy--Wouthuysen Hamiltonian yields the same
result as a covariant calculation.
This is not true for the direct reduction method which will in general lead to
incorrect results even after restoring the gauge invariance property of
the Hamiltonian.
Furthermore, it is shown that an identification of the Z--diagrams of the
usual Dirac representation with the contact graphs of the Foldy--Wouthuysen
representation is incorrect beyond the order of the low--energy theorem.
\end{abstract}

\section{Introduction}
The use of ``effective $2 \times 2$ Hamiltonians'' for relativistic corrections
is one of the main ingredients of many calculations in low-- and
intermediate--energy nuclear physics.
For the electromagnetic interaction of the nucleon such correction terms
beyond the standard Pauli--Hamiltonian \cite{Pauli1,Bjorken} are conventionally
generated through a Foldy--Wouthuysen transformation
\cite{Bjorken,Foldy,Itzykson}.
In some applications the Foldy--Wouthuysen transformation is applied to a
Hamiltonian containing strong interactions as well \cite{Friar,Hyuga,Goeller}.
The algorithm \cite{Bjorken,Foldy,Itzykson} consists of successive applications
of unitary transformations of the wave functions.
It provides a systematic procedure to block--diagonalize the Hamiltonian order
by order in $1/M$, leading to a decoupling of positive-- and negative--energy
states to any desired order in $1/M$.
As the unitary transformations used are in general time--dependent,
the old and the new Hamiltonian are not unitarily equivalent
\cite{Nieto,Goldman}.
The new Hamiltonian rather contains an additional piece involving the unitary
transformation and the time derivative of the inverse transformation.
Nevertheless, such an approach automatically leads to the same result for the
S--matrix as the standard covariant calculation \cite{Fearing}.

In a second approach the matrix element of the interaction Hamiltonian
is evaluated between positive--energy solutions of the free Dirac equation.
The result is then interpreted as the matrix element of some effective
Hamiltonian between two--component Pauli spinors. The so constructed
Hamiltonian is then used in conventional old--fashioned time--ordered
perturbation theory.

In ref.~\cite{Fearing} we have studied for a generic interaction how the
two methods compare and when they are expected to give different results.
We have seen that it is important to investigate processes which
require higher than first--order perturbation theory.
It is the purpose of this work to apply the different methods
to a specific process,
namely low--energy Compton scattering by a proton, in order to make
more quantitative statements.
For several reasons Compton scattering provides an ideal process to test
the approaches mentioned above.
First of all, when expanded in a power series in the frequency $\omega$ of the
photon, the two leading terms of the T--matrix are determined by a low--energy
theorem \cite{GellMann,Low}.
The coefficients of the series are expressed in terms of on--shell properties
of the proton, namely its mass, charge and anomalous magnetic moment.
The derivation of this theorem is based on gauge and relativistic
invariance.
Secondly, it was shown in ref.~\cite{GellMann} that a covariant calculation of
the s-- and u--channel Born terms using the interaction Hamiltonian for a
Dirac particle with an anomalous magnetic moment as proposed by Pauli
\cite{Bjorken,Pauli2} reproduces the low--energy theorem.
Finally, we can address the important question of gauge invariance which
will in general impose some restrictions on the construction of effective
Hamiltonians.

Alternatively we could have considered pion photoproduction at threshold
where two different interactions, namely the strong and the electromagnetic,
enter and a low--energy theorem exists as well
\cite{Kroll,DeBaenst,Naus,Scherer}.
However, the observations for that process are very similar and all the
relevant points can be made using Compton scattering.

Our paper is organized as follows. In the next section  we will define
the interaction Hamiltonians which are used in section 3 to calculate
the T--matrix in second--order perturbation theory. We will discuss the
results to order $\omega^2/M^3$ and comment on gauge invariance.
Finally, in section 4 we will give a short summary.

\section{Definition of the Interaction Hamiltonians}

In this section we will define the different interaction Hamiltonians which
we will use to calculate the corresponding T--matrix element for low--energy
Compton scattering to order $1/M^3$.

\subsection{Relativistic Hamiltonian}

For the relativistic description we will work with the Dirac equation
describing the interaction of a Dirac proton having an anomalous
magnetic moment $(\kappa=1.79)$ with an external electromagnetic field,
\begin{equation}
\label{dirac}
i\frac{\partial \Psi(x)}{\partial t} = (H_0+H_I(t)) \Psi(x),
\end{equation}
where $H_0=\vec{\alpha}\cdot\vec{p}+\beta M$ is the free Dirac Hamiltonian
and $H_I(t)$ is given by \cite{Bjorken,Pauli2}
\begin{equation}
\label{emham}
H_I(t) = \beta (e A^{\mu} \gamma_{\mu} - \frac{e \kappa}{4 M}
         \sigma_{\mu \nu} F^{\mu \nu} ),
\end{equation}
with $F^{\mu \nu} = \partial^{\nu} A^{\mu} - \partial^{\mu} A^{\nu}$.
A calculation involving the model Hamiltonian of eq.~(\ref{emham}) obviously
cannot take {\em all} strong interaction effects into account, such as,
e.~g.~, off--shell effects \cite{Bincer,Nyman,Naus2} or transitions to excited
states.
However, for our purposes this is not required, as we only want
to use the second--order covariant calculation as a reference,
which defines the ``correct'' relativistic result within the
framework of the model Hamiltonian.

\subsection{The Foldy--Wouthuysen Hamiltonian}

The Foldy--Wouthuysen transformation \cite{Bjorken,Foldy,Itzykson} provides a
method
which order by order in $1/M$ decouples the upper from the lower components.
We apply the procedure to the Hamiltonian of eq.~(\ref{dirac}) to obtain a
representation of the $4 \times 4$ Hamiltonian which is block--diagonal to
order
$1/M^3$ \cite{Fearing}.
We define the upper left--hand block of the $4 \times 4$ Foldy--Wouthuysen
Hamiltonian as the effective $2 \times 2$ Hamiltonian, $H^{eff-FW}$, to be
used in our calculation of  Compton scattering by the proton in second--order
perturbation theory.
We keep interaction terms up to second order in the coupling constant $e$
and obtain,
\begin{equation}
\label{hfw}
i \frac{\partial \Psi(x)}{\partial t} = (H_0^{eff-FW}+H^{eff-FW}_1(t)
+H_2^{eff-FW}(t))
\Psi(x),
\end{equation}
with
\begin{eqnarray}
\label{fwhams}
H_0^{eff-FW} & = & M+\frac{\vec{p}\,^2}{2M}-\frac{\vec{p}\,^4}{8M^3}, \nonumber
\\
H_1^{eff-FW}(t) & = & e \Phi - \frac{e}{2M} (\vec{p} \cdot \vec{A}
              + \vec{A} \cdot \vec{p})
              - \frac{e}{2M} (1+\kappa) \vec{\sigma} \cdot \vec{B}
\nonumber \\
& &
-i\frac{e(1+2\kappa)}{8M^2}[\vec{\sigma}\cdot\vec{p},\vec{\sigma}\cdot\vec{E}]
-\frac{e\kappa}{16M^3}\{\vec{\sigma}\cdot\vec{p},\vec{\sigma}\cdot
\dot{\vec{E}}\}
\nonumber \\
& &  +\frac{e}{8M^3} \{\vec{p}\,^2,\{\vec{\sigma}\cdot\vec{p},\vec{\sigma}
\cdot\vec{A}\}\}
+\frac{e\kappa}{16M^3}\{\vec{\sigma}\cdot\vec{p},\{\vec{\sigma}\cdot\vec{p},
\vec{\sigma}\cdot\vec{B}\}\}, \nonumber \\
H_2^{eff-FW}(t) & = & \frac{e^2 \vec{A}^2}{2M} + \frac{e^2}{4 M^2}(1+2 \kappa)
\vec{\sigma}\cdot\vec{E}\times\vec{A} + \frac{e^2}{8M^3}(1+\kappa+\kappa^2)
\vec{E}^2
\nonumber \\
& & -\frac{e^2}{8 M^3} \left (\vec{p}\,^2 \vec{A}^2 + \vec{A}^2 \vec{p}\,^2
    +4 \vec{p} \cdot \vec{A} \, \vec{A} \cdot \vec{p}
    +(\vec{\nabla} \cdot \vec{A})^2 +\vec{B}^2 \right.
\nonumber \\
& & \left. -2 i \vec{p} \cdot \vec{A} \, \vec{\nabla} \cdot \vec{A}
    +2 i \vec{\nabla} \cdot \vec{A} \, \vec{A} \cdot \vec{p}
    +2 \vec{p} \cdot \vec{A} \, \vec{\sigma} \cdot \vec{B}
    +2 \vec{\sigma} \cdot \vec{B} \, \vec{A} \cdot \vec{p} \right )
\nonumber \\
& & - \frac{e^2 \kappa}{8 M^3} \left ( \vec{\sigma} \cdot \vec{p} \,
    \vec{A} \cdot \vec{B} + \vec{A} \cdot \vec{B} \, \vec{\sigma} \cdot \vec{p}
    +\vec{\sigma} \cdot \vec{A} \, \vec{B} \cdot \vec{p}
    +\vec{p} \cdot \vec{B} \, \vec{\sigma} \cdot \vec{A} \right .
\nonumber \\
& & \left. +\vec{A} \cdot \vec{\nabla} \times \vec{B} -\vec{A} \cdot
\dot{\vec{E}}
\right),
\end{eqnarray}
where $\vec{E}=-\vec{\nabla}\Phi-\dot{\vec{A}}$ and
$\vec{B}=\vec{\nabla}\times\vec{A}$ and $\dot{\cal O}$ refers to
$\partial {\cal O}/\partial t$. Note that $\Psi(x)$ in eq.~(\ref{hfw}) denotes
a two--component wave function.

\subsection{Direct Pauli Reduction Hamiltonian}

In many applications a different method is used to obtain an effective
$2 \times 2$ Hamiltonian.
Given a general relativistic interaction, it consists of evaluating the matrix
element of the interaction operator between free positive--energy solutions of
the Dirac equation and reducing it to two--component form.
The result may then be expanded in a power series in $1/M$ and is interpreted
as an effective operator to be used with a free Hamiltonian $H_0^{eff-P}$.
Applying this procedure to the interaction Hamiltonian of eq.~(\ref{emham})
and keeping only terms to order $1/M^3$ yields the following Schr\"odinger
equation
\begin{equation}
\label{schreff}
i \frac{\partial \Psi(x)}{\partial t} = (H_0^{eff-P}+H_I^{eff-P}(t))\Psi(x),
\end{equation}
with $H_0^{eff-P}$ the same as $H_0^{eff-FW}$ in eq.~(\ref{fwhams}) and
\begin{eqnarray}
\label{emhameff}
H_I^{eff-P}(t) & = & e \Phi -
\frac{e}{2M}(\vec{p}\cdot\vec{A}+\vec{A}\cdot\vec{p})
             -\frac{e}{2M}(1+\kappa)\vec{\sigma}\cdot\vec{B} \nonumber\\
& & -\frac{e}{8M^2}[\vec{\sigma}\cdot\vec{p},[\vec{\sigma}\cdot\vec{p},\Phi]]
-i\frac{e\kappa}{4M^2}[\vec{\sigma}\cdot\vec{p},\vec{\sigma}\cdot\vec{E}]
\nonumber\\
& & +\frac{e}{8M^3}\{\vec{p}\,^2,\{\vec{\sigma}\cdot\vec{p},\vec{\sigma}
\cdot\vec{A}\}\}
+\frac{e}{16M^3}[\vec{p}\,^2,[\vec{\sigma}\cdot\vec{p},\vec{\sigma}
\cdot\vec{A}]]
\nonumber \\
& &
+\frac{e\kappa}{16M^3}\{\vec{\sigma}\cdot\vec{p},\{\vec{\sigma}\cdot\vec{p},
\vec{\sigma}\cdot\vec{B}\}\}.
\end{eqnarray}
The $2 \times 2$ effective interaction Hamiltonian $H^{eff-P}_I$ of
eq.~(\ref{emhameff}) may be interpreted as the upper left--hand block
of the transformed operator $H^P_I=T_0 H_I T_0^{\dagger}$, where $T_0$ denotes
the time--independent free Foldy--Wouthuysen transformation
(see e.~g.~eq.~(8) of ref.~\cite{Fearing}).
We emphasize that $H^P_I$ is not block--diagonal, i.~e.~, it will
connect positive-- and negative--energy eigenstates of the
free Foldy--Wouthuysen Hamiltonian.

The above procedure only produces a {\em linear} interaction term
and no contact interactions.
However, it is well--known that a gauge--invariant coupling through a
minimal substitution in $H_0^{eff-P}$ generates linear and quadratic (and
higher--order) terms in $\vec{A}$ as required by gauge invariance.
As such terms are absent in the Hamiltonian of eq.~(\ref{schreff}) it will not
exhibit the correct transformation behavior under a gauge transformation.
Of course, we can cure this problem by introducing additional interaction
terms by hand, such that the resulting Hamiltonian transforms properly.
However, we have to keep in mind that one of our main purposes in
considering Compton scattering is  a {\em critical} examination of
the use of effective Hamiltonians.
In fact, for interactions other than the electromagnetic, there is in
general {\em no} guiding principle which tells us whether or not the simple
matrix element reduction is a reasonable procedure to generate an effective
Hamiltonian.
For that reason we will nevertheless use the Hamiltonian of
eq.~(\ref{emhameff}) (being aware that this will lead to unacceptable results)
and will postpone the question of gauge invariance until we discuss
our results.

\section{Calculation of the T--Matrix Element}
In this section we will calculate the T--matrix element for low--energy
Compton scattering in second--order perturbation theory using the different
interaction Hamiltonians of section 2, eq.~(\ref{emham}), (\ref{fwhams})
and (\ref{emhameff}), and compare the results.

The kinematical variables and polarization vectors for Compton scattering
are defined in figure 1.
The calculations are performed in the laboratory frame, where the energies
$\omega$ and $\omega'$ of the initial and final photon are related by
\begin{equation}
\label{energies}
\omega' = \omega \left (1+\frac{\omega}{M}(1-\cos(\theta)) \right)^{-1}
        = \omega \left (1-\frac{\omega}{M}(1-\cos(\theta))
          + [\frac{\omega^2}{M^2}] \right).
\end{equation}
In eq.~(\ref{energies}) $\theta$ denotes the angle between the momenta of the
initial and the final photon in the laboratory system and $[\omega^2/M^2]$
means ``of order of $\omega^2/M^2$''.
Furthermore, we make use of the radiation gauge,
$\Phi=0,\vec{\nabla} \cdot \vec{A}=0$.
We express the result as an operator which still has to be evaluated
between two--component Pauli spinors of the proton.
Any calculation of the T--matrix element\footnote{The
T--matrix element $t_{fi}$ is related to the invariant amplitude ${\cal M}$
of ref.~\cite{Bjorken} through $t_{fi}=-i (M/\sqrt{E_i E_f}) {\cal M}$.}
to order $1/M^3$ may be written in the following form
\cite{Low,Petrun'kin}
\begin{eqnarray}
\label{tpar}
t_{fi} & = &  \vec{\epsilon} \cdot \vec{\epsilon'} A_1
        + i \vec{\sigma} \cdot \vec{\epsilon'} \times \vec{\epsilon} A_2
        + \hat{k'} \times \vec{\epsilon'} \cdot \hat{k} \times \vec{\epsilon}
A_3
        + i \vec{\sigma} \cdot (\hat{k'} \times \vec{\epsilon'})
        \times (\hat{k} \times \vec{\epsilon}) A_4
\nonumber \\
  & &   + i \hat{k} \cdot \vec{\epsilon'} \, \vec{\sigma} \cdot \hat{k}
        \times \vec{\epsilon} A_5
        + i \hat{k'} \cdot \vec{\epsilon} \, \vec{\sigma} \cdot \hat{k'}
        \times \vec{\epsilon'} A_6
\nonumber \\ & &
        + i \hat{k} \cdot \vec{\epsilon'} \, \vec{\sigma} \cdot \hat{k'}
        \times \vec{\epsilon} A_7
        + i \hat{k'} \cdot \vec{\epsilon} \, \vec{\sigma} \cdot \hat{k}
        \times \vec{\epsilon'} A_8,
\end{eqnarray}
where $\hat{k},\vec{\epsilon}\, (\hat{k'},\vec{\epsilon'})$ refer to the
direction and the polarization of the initial (final) photon,
respectively\footnote{It is possible to construct a generalized Hamiltonian
which parameterizes the T--matrix including all strong interaction effects
to order $\omega^2/M^3$ \cite{Petrun'kin,L'vov}. Such an effective Hamiltonian
involves
the Compton polarizabilities $\bar{\alpha}$ and $\bar{\beta}$.}.
In the following we will calculate the amplitudes $A_i$ as predicted to
second--order perturbation theory using the different Hamiltonians.

\subsection{Covariant Calculation}
In the covariant calculation the invariant amplitude to second--order
perturbation theory using the Hamiltonian of eq.~(\ref{emham}) reads
\begin{eqnarray}
\label{covam}
{\cal M} & = & -i e^2 \bar{u}(p_f)\left(\not\epsilon'
-\frac{\kappa}{2 M}\not k'\not\epsilon'\right)
\frac{1}{\not p_i+\not k-M+i\delta}
\left(\not\epsilon+\frac{\kappa}{2 M}\not k\not\epsilon\right)u(p_i)
\nonumber\\
& & + (\epsilon\leftrightarrow\epsilon',k\leftrightarrow-k'),
\end{eqnarray}
where the expression for the u--channel diagram may be obtained from the
s--channel diagram through crossing symmetry \cite{Bjorken} as
indicated in eq.~(\ref{covam}). In eq.~(\ref{covam}) we have introduced
$-i\delta$ instead of the standard $-i\epsilon$ as the small imaginary mass
required by the Feynman--St\"uckelberg boundary condition
in order to avoid confusion with the polarization vector $\epsilon^{\mu}$.

For our discussion it turns out to be useful to split the Feynman propagator
of the nucleon, $S_F(p)$, into its positive-- and negative--frequency
contribution,
\begin{eqnarray}
\label{feynmanpropagator}
S_F(p) & = & \frac{1}{\not p - m + i \delta} \nonumber \\
       & = &  S_F^{(+)}(p) + S_F^{(-)}(p) \nonumber \\
       & = & \frac{E \gamma^0 - \vec{p} \cdot
             \vec{\gamma} + M}{2 E (p^0-E+i\delta)}
           + \frac{E \gamma^0 + \vec{p} \cdot
             \vec{\gamma} - M}{2 E (p^0+E-i\delta)},
\end{eqnarray}
with $E=\sqrt{\vec{p}\,^2 + M^2}$. Such a separation allows one to identify
Z--diagrams and ordinary diagrams of old--fashioned time--ordered
perturbation theory (see e.~g.~ref.~\cite{Weinberg} and figure 2). It should,
however, be noted that, although the individual diagrams of covariant
perturbation theory are Lorentz scalars, this is not separately true
for their positive-- and negative--frequency contributions \cite{Weinberg}.
In other words, these contributions are frame--dependent quantities.
Furthermore, as was pointed out in ref.~\cite{Jennings}, the notion of
Z--diagrams depends on the interaction as well as the representation used.
Here, we will refer to Z--diagrams (ordinary diagrams) as those resulting
from the negative-- (positive--) frequency part of the propagator of
eq.~(\ref{feynmanpropagator}) when using the Hamiltonian of eq.~(\ref{emham}).

After inserting eq.~(\ref{feynmanpropagator}) into eq.~(\ref{covam}) we
expand the resulting expression in $1/M$ and bring it into the form of
eq.~(\ref{tpar}).
The predictions of this straightforward but tedious calculation for the
amplitudes $A_i$ separated into their forward and backward contributions
are tabulated in table 1 and 2.

\subsection{Foldy--Wouthuysen Calculation}

The calculation involving the effective Foldy--Wouthuysen Hamiltonian
of eq.~(\ref{fwhams}) is performed in second--order time--ordered perturbation
theory
and the contribution to the S--matrix of second order in the coupling
constant reads
\begin{eqnarray}
\label{soptfw}
S_{fi} & = & -2 \pi i \delta (E_f+\omega'-E_i-\omega) \Bigg \{
             <\Phi^{FW}_{0f}|H^{eff-FW}_2|\Phi^{FW}_{0i}> \nonumber \\
& & + \sum_{spins} \int d^3p \frac{<\Phi^{FW}_{0f}|H^{eff-FW}_{1 em}|
\Phi^{FW}_{0\vec{p}}><\Phi^{FW}_{0\vec{p}}|H^{eff-FW}_{1 abs}|
\Phi^{FW}_{0i}>}{E_f+\omega'-E_{\vec{p}} + i\delta} \nonumber \\
& & + \sum_{spins} \int d^3p \frac{<\Phi^{FW}_{0f}|H^{eff-FW}_{1 abs}|
\Phi^{FW}_{0\vec{p}}><\Phi^{FW}_{0\vec{p}}|H^{eff-FW}_{1 em}|
\Phi^{FW}_{0i}>}{E_f-\omega-E_{\vec{p}} + i\delta}\Bigg\}. \nonumber \\
& &
\end{eqnarray}
In arriving at eq.~(\ref{soptfw}) we have assumed a harmonic time dependence
for the interaction, i.~e.~, $\exp(-i\omega t)$ and $\exp(i\omega't)$
for the initial and final photon appearing in the interaction Hamiltonian
of eq.~(\ref{fwhams}), respectively. Furthermore, $H^{eff-FW}_{1 em}$ and
$H^{eff-FW}_{1 abs}$ refer to the emission and absorption of a
photon, respectively, and it is understood that the part of $H^{eff-FW}_2$
which at the same time creates and annihilates a photon is used.
The Hamiltonian $H_2^{eff-FW}$ gives rise to a contact term, whereas
$H_1^{eff-FW}$ contributes ``quadratically'' in second--order perturbation
theory (see figure 3).
In Coulomb gauge, the leading--order term of $H^{eff-FW}_1$ is of order $1/M$.
As we are interested in the T--matrix element to order $1/M^3$,
we need the linear interaction Hamiltonian to order $1/M^2$
only\footnote{If one wants to check gauge invariance
to order $1/M^3$ one has to retain terms to order $1/M^3$ in
$H^{eff-FW}_1$, since the leading--order term in $H^{eff-FW}_1$ is proportional
to
$\Phi$.}.
Establishing the graphical rules corresponding to the interaction Hamiltonians
of eq.~(\ref{fwhams}) is straightforward, but the expressions are lengthy.
We only report the predictions for the amplitudes $A_i$ in table 1 and 2,
separated according to their origin as contact or non--contact terms.

\subsection{Calculation Involving the Effective Hamiltonian
$H^{eff-P}_I$ and Gauge Invariance}
Finally, a second--order calculation involving the effective Hamiltonian of
eq.~(\ref{emhameff}), $H^{eff-P}_I$, gives rise to
\begin{eqnarray}
\label{soptp}
S_{fi} & = & -2 \pi i \delta (E_f+\omega'-E_i-\omega) \times \nonumber \\
& & \Bigg \{
\sum_{spins} \int d^3p \frac{<\Phi^{FW}_{0f}|H^{eff-P}_{I em}|
\Phi^{FW}_{0\vec{p}}><\Phi^{FW}_{0\vec{p}}|H^{eff-P}_{I abs}|
\Phi^{FW}_{0i}>}{E_f+\omega'-E_{\vec{p}} + i\delta} \nonumber \\
& & + \sum_{spins} \int d^3p \frac{<\Phi^{FW}_{0f}|H^{eff-P}_{I abs}|
\Phi^{FW}_{0\vec{p}}><\Phi^{FW}_{0\vec{p}}|H^{eff-P}_{I em}|
\Phi^{FW}_{0i}>}{E_f-\omega-E_{\vec{p}} + i\delta}\Bigg\}. \nonumber \\
& &
\end{eqnarray}
As we have seen in ref.~\cite{Fearing} such a calculation amounts to
neglecting the negative--frequency part of the covariant calculation.
Thus we already know the results for the $A_i$, as we have split
the covariant calculation into its forward and backward part.

Approximations, such as the neglect of the negative--energy contributions, will
in general be in conflict with gauge invariance.
Thus we have to address the gauge invariance property of the equation of
motion\footnote{Although we only write down the potentials $\Phi$ and $\vec{A}$
as the argument of the Hamiltonian, it is understood that it may as well
depend on derivatives of the potentials.}
\begin{equation}
\label{eqmo}
i\frac{\partial\Psi(x)}{\partial t} = H(\Phi,\vec{A})\Psi(x).
\end{equation}
Gauge invariance means that
\begin{equation}
\label{psip}
\Psi'(x)=\exp(-ie\chi(x))\Psi(x)
\end{equation}
is a solution of
\begin{equation}
\label{eqmop}
i\frac{\partial\Psi'(x)}{\partial t} = H(\Phi+\dot{\chi},
\vec{A}-\vec{\nabla}\chi)\Psi'(x),
\end{equation}
provided $\Psi(x)$ is a solution of eq.~(\ref{eqmo}) \cite{Messiah}. From this
we may derive the following constraint for the form of the Hamiltonian
\begin{equation}
\label{condition}
H(\Phi+\dot{\chi},\vec{A}-\vec{\nabla}\chi)
- \exp(-i e \chi)H(\Phi,\vec{A})\exp(i e \chi)-e\dot{\chi} = 0.
\end{equation}
The effective Foldy--Wouthuysen Hamiltonian
of eq.~(\ref{hfw}) satisfies the gauge invariance property to order
$e^2$ whereas the effective Hamiltonian of eq.~(\ref{schreff})
is not gauge--invariant even to first order in $e$.
Suppose we denoted the effective Hamiltonian of eq.~(\ref{schreff})
by $H_A$ and we wanted to add a Hamiltonian $H_B$ such that
the sum satisfied the condition of eq.~(\ref{condition}).
Inserting $H_A+H_B$ into eq.~(\ref{condition}) it is easily seen that
the condition for $H_B$ reads
\begin{eqnarray}
\label{newcondition}
\lefteqn{\exp(-i e \chi)H_B(\Phi,\vec{A})\exp(i e \chi)
-H_B(\Phi+\dot{\chi},\vec{A}-\vec{\nabla}\chi)=} \nonumber \\ &&
H_A(\Phi+\dot{\chi},\vec{A}-\vec{\nabla}\chi)
-\exp(-i e \chi)H_A(\Phi,\vec{A})\exp(i e \chi)-e\dot{\chi}.
\end{eqnarray}
{}From eq.~(\ref{newcondition}) it is seen that $H_B$ cannot be determined
uniquely as one may always add another term $H_C$ which separately satisfies
\begin{equation}
\label{conditionc}
\exp(-i e \chi)H_C(\Phi,\vec{A})\exp(i e \chi)
-H_C(\Phi+\dot{\chi},\vec{A}-\vec{\nabla}\chi) = 0.
\end{equation}

Let us for convenience consider eq.~(\ref{schreff}) to order
$1/M^2$ only, as this is sufficient for the point we want to make.
Inserting $H_A=H_0^{eff-P}+H^{eff-P}_I$ into eq.~(\ref{newcondition}) we obtain
the following
constraint for $H_B$
\begin{eqnarray}
\label{cond}
\lefteqn{\exp(-i e \chi)H_B(\Phi,\vec{A})\exp(i e \chi)-
H_B(\Phi+\dot{\chi},\vec{A}-\vec{\nabla}\chi) =}
\nonumber \\
& & -\frac{e}{8M^2}[\vec{\sigma}\cdot\vec{p},[\vec{\sigma}\cdot\vec{p},
\dot{\chi}]]+\frac{e^2}{2M}(2\vec{A}\cdot\vec{\nabla}\chi
-(\vec{\nabla}\chi)^2) \nonumber \\
& & -\frac{ie^2}{8M^2}[\chi,[\vec{\sigma}\cdot\vec{p},
[\vec{\sigma}\cdot\vec{p},\Phi]]]
+\frac{e^2\kappa}{4M^2}[\chi,[\vec{\sigma}\cdot\vec{p},
\vec{\sigma}\cdot\vec{E}]] + [e^3].
\end{eqnarray}
{}From the right--hand side of eq.~(\ref{cond}) we can now see
that $H_0^{eff-P}+H^{eff-P}_I$ does not satisfy the constraint
of eq.~(\ref{condition}) even to leading order in $e$.
The introduction of e.~g.
\begin{equation}
\label{hb}
H_B(\Phi,\vec{A})=\frac{e}{8M^2}[\vec{\sigma}\cdot\vec{p},
[\vec{\sigma}\cdot\vec{p},\Phi]]+\frac{e^2\vec{A}^2}{2M}+
\frac{i e^2\kappa}{4M^2}[\vec{\sigma}\cdot\vec{A},\vec{\sigma}\cdot\vec{E}]
\end{equation}
makes the sum $H_A+H_B$ transform according to eq.~(\ref{condition}) to
order $e^2$ and $1/M^2$. The second term of eq.~(\ref{hb})
``completes'' the minimal substitution in the kinetic--energy
term of eq.~(\ref{schreff}) to the order we are considering here.
The other two terms are clearly not unique and a comparison
with the Foldy--Wouthuysen Hamiltonian to the same order, namely
$1/M^2$, shows that the two Hamiltonians are in fact different.

\subsection{Results and Discussion}

In table 1 we show the results for the amplitudes $A_i$ to the order of
the low--energy theorem, namely $[1/M]$ and $[\omega/M^2]$.
We find that the negative--frequency contribution of the
covariant calculation (Z--diagrams) yields exactly the same result as the
contact interaction of the Foldy--Wouthuysen transformation.
Similarly, we have as well a correspondence between the positive--frequency
part
(ordinary diagrams) on the one hand and the non--contact terms
of the FW calculation on the other hand.

The {\em naive} -- but within the framework of nonrelativistic
second--order perturbation theory
consistent -- calculation involving the effective Hamiltonian of
eq.~(\ref{emhameff}) is identical with the positive--frequency
contribution of the covariant calculation and leads to completely
unacceptable results.
Even the leading--order term, the so--called Thomson limit, is not reproduced.
Inspection of table 1 tells us that the neglect of the negative--frequency
contribution is responsible for such behavior.
Such an approximation will spoil the gauge invariance property of the model.
However, it is well known that the low--energy theorem for the Thomson
limit is based on gauge invariance.
For that reason we constructed an additional interaction term
to make the total Hamiltonian gauge--invariant.
This construction naturally involved the $e^2\vec{A}^2/2M$ term which then
generated the correct Thomson limit.
However, the other terms are by no means unique.
As we can see from table 1 even including the additional Hamiltonian $H_B$ of
eq.~(\ref{hb}) does not reproduce the correct low--energy theorem
prediction beyond the leading--order term, although the Hamiltonian
used is gauge--invariant to order $e^2$.
It was pointed out by
Low \cite{Low} that the determination of the
terms beyond the the Thomson limit require more input than gauge
invariance alone. Our example clearly is in agreement with this
observation.
In other words, enforcing gauge invariance is required to obtain the
Thomson limit, but it does not predict the next--to--leading--order
term of the low--energy theorem.

The Compton scattering by a proton firstly demonstrates how misleading
it may be to use effective Hamiltonians obtained from a two--component
reduction of a relativistic matrix element.
The failure to reproduce the contributions of the Z--diagrams
originates from the fact that the $2 \times 2$ effective Hamiltonian
$H^{eff-P}_I$ is generated from a $4 \times 4$ Hamiltonian $H^P_I$
which is {\em not} block--diagonal.
Use of $H^{eff-P}_I$ then amounts to a truncation of the Hilbert space,
namely the neglect of negative--energy states in this approach, which will
in general lead to unreliable results.
In contrast, the $4 \times 4$ Foldy--Wouthuysen Hamiltonian $H^{FW}$ is
block--diagonal and hence using the $2 \times 2$ Hamiltonian $H^{eff-FW}$
does not make a difference,
as $H^{FW}$ does not connect positive--
and negative--energy states in the FW representation.

Furthermore, we have seen that gauge invariance alone is in general not
sufficient to overcome this shortcoming, i.~e.~, to determine
more than the leading--order term.
For a similar discussion of the second point within the framework of
a quantum field theoretical model see ref.~\cite{Bos}.

The terms of order $\omega^2/M^3$ are beyond the prediction of the
low--energy theorem and hence are model--dependent.
The results of the covariant calculation and the Foldy--Wouthuysen
calculation for these terms are listed in table 2.
Once again we see that the {\em total} result of both calculations
is the same at this order in agreement with our observations in
ref.~\cite{Fearing}.
However, from table 1 one might have concluded that the correspondence
between Z--diagrams and contact interactions and equally between
ordinary diagrams and non--contact terms holds true for
higher orders in $1/M$ as well.
It can now be seen that for some of the amplitudes $A_i$
this is no longer true.
In order to be specific, we find that the Z--diagram contribution to the
amplitudes $A_1$, $A_5$ and $A_6$ differs from the contact terms.

As in table 1 the result of the effective Hamiltonian is given by
the positive--frequency part of the covariant calculation.
We have not made the effort of forcing the effective Hamiltonian
to be gauge--invariant to order $e^2$ and $1/M^3$ since the result
will not be unique anyway.

\section{Summary and Conclusions}

In this work we considered Compton scattering by the proton within three
different approaches.
Firstly, we performed a covariant calculation to second--order perturbation
theory.
In order to identify Z--diagrams and ordinary diagrams in the covariant
approach, we split the Feynman propagator into its positive-- and
negative--frequency contribution.
We expanded the so obtained T--matrix element to order $\omega^2/M^3$ and
used the result as a reference for the nonrelativistic calculations with
effective Hamiltonians.

The calculation involving the Foldy--Wouthuysen Hamiltonian correctly
reproduced the reference result of the covariant approach.
This was expected as the $4 \times 4$ Foldy--Wouthuysen Hamiltonian
is block--diagonal and thus does not couple between positive-- and
negative--energy states. The neglect of the negative--energy states in going
from the full S--matrix to the effective one
has no consequence and hence the full (and as a reference correct)
S--matrix is still reproduced.
However, we saw that an identification of the Z--diagrams with the contact
interaction of the Foldy--Wouthuysen Hamiltonian is {\em not} correct beyond
the order of the low--energy theorem.

Finally, the effective Hamiltonian $H^{eff-P}_I$ obtained from a direct
reduction of the relativistic interaction Hamiltonian between positive--energy
solutions of the free Dirac equation led to incorrect results.
The reason is that $H^{eff-P}_I$ is derived as the upper left--hand
block from a $4 \times 4$ interaction Hamiltonian which is not
block--diagonal.
In this case, the neglect of the negative--frequency solutions
makes a considerable difference. Even the leading--order term
is not reproduced.
This may be cured in part by enforcing the gauge invariance property
of the corresponding Hamiltonian.
However, as we saw, such a procedure is not unique, and thus there will
always be some degree of arbitrariness involved, when enforcing gauge
invariance.

\section{Acknowledgements}
This work was supported in part by a grant from the Natural Sciences and
Engineering Research Council of Canada. The authors would like to thank
B.~K.~Jennings for useful discussions.

\newpage
\begin{table}
\begin{tabular}{|c|| c | c || c || c | c |}
\hline
&\multicolumn{2}{c||}{covariant calculation}& &\multicolumn{2}{|c|}{FW
calculation
in s.~o.~p.~t.} \\
\hline
amplitude & negative & positive & $H_B$ & contact & non--contact \\
\hline
\hline
& & & & & \\
$A_1$ & $\frac{1}{M}$ & 0 & $\frac{1}{M}$ & $\frac{1}{M}$ & 0 \\
& & & & & \\
\hline
& & & & & \\
$A_2$ & $-\frac{(\omega+\omega')(1+2\kappa)}{4M^2}$ & 0 &
$-\frac{(\omega+\omega')\kappa}{2M^2}$ &
$-\frac{(\omega+\omega')(1+2\kappa)}{4M^2}$ & 0 \\
& & & & & \\
\hline
& & & & & \\
$A_4$ & 0 & $\frac{(\omega+\omega')(1+\kappa)^2}{4M^2}$ & 0 & 0 &
$\frac{(\omega+\omega')(1+\kappa)^2}{4M^2}$ \\
& & & & & \\
\hline
& & & & & \\
$A_5$ & 0 & $ \frac{\omega (1+\kappa)}{2M^2}$ & 0
& 0 & $ \frac{\omega (1+\kappa)}{2M^2}$ \\
& & & & & \\
\hline
& & & & & \\
$A_6$ & 0 & $-\frac{\omega' (1+\kappa)}{2M^2}$ & 0
& 0 & $-\frac{\omega' (1+\kappa)}{2M^2}$ \\
& & & & & \\
\hline
\end{tabular}
\\
\\
{\bf Table 1: Compton Scattering amplitudes}\\
Contribution to the amplitudes $A_i$ to the order of the low--energy
theorem in units of $e^2$.
The amplitudes $A_3$, $A_7$ and $A_8$ are zero at this order.
The second and third column refer to the negative-- and
positive--frequency contribution (Z--diagrams and ordinary diagrams) of the
covariant calculation at second--order perturbation theory.
The positive--frequency part of the covariant calculation is identical
with the result of the effective Hamiltonian resulting from a
direct Pauli reduction.
The fourth column contains the additional part due to $H_B$
in order to make the effective Hamiltonian gauge--invariant.
The fifth and sixth column refer to the contact and non--contact terms
of the Foldy--Wouthuysen calculation in second--order perturbation
theory.
\end{table}

\newpage

\begin{table}
\begin{tabular}{|c|| c | c || c | c |}
\hline
&\multicolumn{2}{c||}{covariant calculation}& \multicolumn{2}{|c|}{FW
calculation
in s.~o.~p.~t.} \\
\hline
amplitude & negative & positive & contact & non--contact \\
\hline
\hline
& & & & \\
$A_1$ & $-\frac{\omega^2(1-\cos(\theta))}{2M^3}$
      & $-\frac{\omega^2\kappa(1+\kappa)}{2M^3}$
      & $-\frac{\omega^2(1-\cos(\theta))}{2M^3}$
      & $-\frac{\omega^2(1+\kappa)(1+2\kappa)}{4M^3}$ \\
      & $+\frac{\omega^2\kappa^2}{4M^3}$ &
      & $+\frac{\omega^2(1+\kappa+\kappa^2)}{4M^3}$ &  \\
& & & & \\
\hline
& & & & \\
$A_3$ & $-\frac{\omega^2}{4M^3}$
      & $\frac{\omega^2(1+\kappa)^2\cos(\theta)}{4M^3}$
      & $ -\frac{\omega^2}{4M^3}$
      & $\frac{\omega^2(1+\kappa)^2\cos(\theta)}{4M^3}$\\
      & & & & \\
\hline
& & & & \\
$A_5$ & $-\frac{\omega^2}{4M^3}$
      & $\frac{\omega^2(1+\kappa+\kappa^2)}{4M^3}$
      & $-\frac{\omega^2}{4M^3}-\frac{\omega^2 \kappa}{8M^3}$
      & $\frac{\omega^2(2+3\kappa+2\kappa^2)}{8M^3}$ \\
& & & & \\
\hline
& & & & \\
$A_6$ & $-\frac{\omega^2}{4M^3}$
      & $\frac{\omega^2(1+\kappa+\kappa^2)}{4M^3}$
      & $-\frac{\omega^2}{4M^3}-\frac{\omega^2 \kappa}{8M^3}$
      & $\frac{\omega^2(2+3\kappa+2\kappa^2)}{8M^3}$ \\
& & & & \\
\hline
& & & & \\
$A_7$ & $-\frac{\omega^2\kappa}{4M^3}$ & 0
      & $-\frac{\omega^2\kappa}{4M^3}$ & 0 \\
& & & & \\
\hline
& & & & \\
$A_8$ & $-\frac{\omega^2\kappa}{4M^3}$ & 0
      & $-\frac{\omega^2\kappa}{4M^3}$ & 0 \\
& & & & \\
\hline
\end{tabular}
\\
\\
{\bf Table2: Compton Scattering amplitudes\\}
Contributions to the amplitudes $A_i$ beyond the low--energy theorem
to order $\omega^2/M^3$ in units of $e^2$. The amplitudes $A_2$ and
$A_4$ are zero at this order.
Note that we have not constructed a Hamiltonian $H_B$.
\end{table}

\end{document}